\documentclass[twocolumn]{emulateapj}
\usepackage{graphicx,subfigure} 
\usepackage{amsmath} 
\usepackage{hyperref}
\usepackage{color}
\def\arcsec{\hbox{$^{\prime\prime}$}}
\def\deg{\hbox{$^\circ$}}

\newcommand{\init}{\hspace{0.7mm}}

\begin{document}
\title{
Identification of 1.4 Million AGNs in the Mid-Infrared using \textit{WISE} Data
}

\author{
N.\init J.\init Secrest\altaffilmark{1}, R.\init P.\init Dudik\altaffilmark{1}, B.\init N.\init Dorland\altaffilmark{1}, N.\init Zacharias\altaffilmark{1}, V.\init Makarov\altaffilmark{1}, A.\init Fey\altaffilmark{1}, J.\init Frouard\altaffilmark{1}, \& C.\init Finch\altaffilmark{1}
 }

\altaffiltext{1}{U.S.\ Naval Observatory, 3450 Massachusetts Avenue NW, Washington, DC 20392, USA}

\begin{abstract}

We present an all-sky sample of $\approx1.4$~million AGNs meeting a two color infrared photometric selection criteria for AGNs as applied to sources from the \textit{Wide-Field Infrared Survey Explorer} final catalog release (AllWISE).  We assess the spatial distribution and optical properties of our sample and find that the results are consistent with expectations for AGNs.  These sources have a mean density of $\approx 38$ AGNs per square degree on the sky, and their apparent magnitude distribution peaks at $g \approx 20$, extending to objects as faint as $g\approx26$.  We test the AGN selection criteria against a large sample of optically-identified stars and determine the ``leakage'' (that is, the probability that a star detected in an optical survey will be misidentified as a QSO in our sample) rate to be $\le4.0\times10^{-5}$.  We conclude that our sample contains almost no optically-identified stars ($\leq0.041\%$), making this sample highly promising for future celestial reference frame work by significantly increasing the number of all-sky, compact extragalactic objects.  We further compare our sample to catalogs of known AGNs/QSOs and find a completeness value of $\gtrsim84\%$ (that is, the probability of correctly identifying a known AGN/QSO is at least $84\%$) for AGNs brighter than a limiting magnitude of $R\lesssim19$.  Our sample includes approximately 1.1~million previously uncatalogued AGNs.

\end{abstract}


\section{Introduction}

The International Celestial Reference Frame (ICRF) is the realization, at radio wavelengths, of the International Celestial Reference System (ICRS), the solar-barycentric, quasi-inertial fundamental reference system adopted by the International Astronomical Union \citep{Arias+95}.  The second realization, ICRF2, consists of 3,414 compact radio objects (i.e., QSOs), of which 295 are ``defining sources'' \citep{Fey+15}.  These QSOs can be nearly ideal reference frame objects, as they present no significant parallax or proper motion and, when properly selected, they have minimal spatial structure or variability.  Using Very Long Baseline Interferometry (VLBI) techniques, ICRF2 defining source position errors have an estimated noise floor of 40~$\mu$as.  Plans for the third realization of the ICRF (ICRF3), to be released in the 2018 timeframe, focus on further densifying the source catalog and improving spatial uniformity (especially at negative declinations), improving the astrometric accuracy of the non-defining sources to bring them close to the accuracy of the defining sources, and extending of the ICRF to higher frequencies to reduce the effects of source structure on position \citep{Jacobs+14}.

Much of the motivation for this improvement to the radio reference frame is driven by the European Space Agency's (ESA) \textit{Gaia} mission, launched in 2013.  \textit{Gaia} is a space-based, astrometric, photometric and radial velocity all-sky survey at optical wavelengths.  Over the next few years, \textit{Gaia} will deliver astrometric catalogs that are expected to be adopted as the next optical instantiation of the fundamental reference frame.  Unlike its predecessor, \textit{Hipparcos}, which was limited to observing stars only, \textit{Gaia}, with its limiting magnitude of $V \approx 20$, will directly observe hundreds of thousands of extragalactic objects.  By tying together the  positions of reference objects (i.e., QSOs) observed in both radio and optical, the \textit{Gaia} reference frame will be brought into alignment and rotationally stabilized with respect to the radio reference frame. 

One critical limitation in aligning optical and radio reference frames is the problem of discrepancies (or ``offsets'') in position between radio and optical measurements.  Offsets can be due to a variety of underlying physical differences between the emission mechanisms of AGNs in the optical and in the radio.  First, for AGN-dominated galaxies, optical emission is thought to originate from the compact accretion disk surrounding the supermassive black hole (SMBH), while radio emission can be either compact or extended, depending on the orientation of the jet with respect to the observer.  Second, for non AGN-dominated galaxies, an optical centroid can be shifted relative to the radio position because of contamination by the host galaxy.  Depending on the distance to the source, the optical position can be significantly different from the position of a compact radio core.  Third, variability in either the jet or the accretion disk can cause apparent changes in the overall position of the AGN in the optical or the radio over time, making the AGN an unreliable tie source for use in defining the reference frame.  

Position offsets as a function of wavelength in this class of sources have been extensively studied, beginning with ~\citet{daSilvaNeto+02} on the correlation between offsets and X-band (8.4~GHz) structure in ICRF sources, and extending to large systematic analyses of radio sources overlapping with the \textit{Sloan Digital Sky Survey}~(SDSS)~\citep{OroszFrey+13}.  Recent optical observations of ICRF sources found offsets between optical and radio positions at the 10~mas level, leaving the authors to conclude that unknown effects, most likely having to do with underlying astrophysical phenomena, induce this real offset between radio and optical positions.  Such an offset would introduce a fundamental limit of about 0.5 mas to the accuracy of the alignment between Gaia and the ICRF \citep{Zacharias+14}, and would be exacerbated should the sources be time varying in nature.  These offsets can potentially be minimized by selecting reference frame objects that minimize problematic features, such as photometric variability or optical/radio structure.  

It is thus crucial to identify and characterize as many AGNs/QSOs as possible in order to maximize the number of reference frame tie objects.   Until recently,  the number of known QSOs was quite limited, of order a few tens of thousands over the entire sky.  The SDSS-DR12 Quasar Catalog (DR12Q), using data from the Baryon Oscillation Spectroscopic Survey~\citep[BOSS;][]{Dawson+13} and photometrically-selected QSO candidates from SDSS, contains 297,301 spectroscopically confirmed QSOs over approximately 9,200 deg$^2$ ($\approx 22\%$ of the sky; Isabelle P\^aris 2015, private communication).  \textit{Gaia} is expected to observe approximately half a million QSOs over the course of its mission \citep{Claeskens+06,Mignard+12}.  A large, all-sky, coherent catalog of well-defined zero-parallax, zero-proper motion sources that includes but also extends beyond the network of ICRF sources would permit an extensive study of all physical offsets associated with radio and optical reference frames, object positional stability, and object variability, using ground-based resources currently available, such as the United States Naval Observatory (USNO) Robotic Astrometric Telescope~\citep[URAT,][]{Zacharias05,Zacharias+15}, and the Panoramic Survey Telescope and Rapid Response System~\citep[Pan-STARRS,][]{Kaiser+10,Morganson+14}.

In order to identify and characterize as many AGNs/QSOs uniformly selected across the sky as possible, we apply a two-color AGN selection criterion to the \textit{Wide-Field Infrared Survey Explorer}~\citep[\textit{WISE},][]{Wright+10} database.  We describe our methodology in the following sections, and discuss the properties of the resultant sample and the reliability of the selection criteria we have chosen for detecting AGNs.  Our resultant sample contains $\approx1.4$~million AGNs, of which $\approx1.1$~million are previously uncatalogued, and most are compact (subtending $\lesssim1\arcsec$ - $2\arcsec$).  The goal of generating this sample is to provide a large sample of point-like extragalactic objects with minimal stellar contamination for the purpose of supporting future photometric, astrometric, and variability studies; maintenance and improvements of the celestial reference frame; and general astrophysical and cosmological uses.  This paper is outlined as follows:  In \S\ref{sec:AllWISEcatalog} we review the \textit{WISE} mission and the AllWISE source catalog and its properties; in \S\ref{sec:AGNselection} we detail the AGN/QSO selection criteria we use; finally in \S\ref{sec:Results} we discuss the resultant sample, as well as some of the properties of the sample sources derived using the \textit{Sloan Digital Sky Survey}, data release 12 \citep[SDSS-DR12,][]{Alam+15}.

\section{AllWISE Catalog}
\label{sec:AllWISEcatalog}
The \textit{WISE} survey is an all-sky mid-IR survey at 3.4, 4.6, 12, and 22 $\mu$m ($W1$, $W2$, $W3$, and $W4$, respectively), conducted between January 7 and August 6, 2010, during the cryogenic mission phase, and first made available to the public on April 14, 2011. \textit{WISE} has an angular resolution of $6.1''$, $6.4''$, $6.5''$, and $12.0''$ in its four bands.  The AllWISE data release, which we use for this work, incorporates data from the \textit{WISE} Full Cryogenic, 3-Band Cryo, and NEOWISE Post-Cryo survey~\citep{Mainzer+14} phases, which were coadded to achieve a depth of coverage $\approx0.4$ magnitudes deeper than previous data releases.\footnote{The increase in depth is primarily due to the additional coverage in the $W1$ and $W2$ bands during the Post-Cryo survey phase, although photometry in all four bands has been improved due in part to better background estimation.} AllWISE contains positions, apparent motions, magnitudes, and PSF-profile fit information for almost 748 million objects.    Astrometric calibration of sources in the \textit{WISE} catalog was done by correlation with bright stars from the 2MASS point source catalog, and the astrometric accuracy for sources in the AllWISE release was further improved by taking into account the proper motions of these reference stars, taken from the fourth USNO CCD Astrograph Catalog~\citep[UCAC4,][]{Zacharias+13}.  A comparison with ICRF sources shows that AllWISE Catalog sources between $8<W1<12$~mag have positional accuracies to within 50~mas, and half of these sources have positional accuracies to within 20~mas. For more details on the \textit{WISE} mission, see~\citet{Wright+10}.

\section{AGN Selection}
\label{sec:AGNselection}
There are numerous mid-IR color selection criteria in the literature, such as the \textit{Spitzer} two-color criteria in \citet{Lacy+04}, \citet{Stern+05}, \citet{Lacy+07}, and \citet{Donley+12}; the \textit{WISE} two-color criteria in \citet{Jarrett+11}, the \textit{WISE} one-color criteria of \citet{Stern+12} and \citet{Assef+13}, and the \textit{WISE} two-color criteria of \citet{Mateos+12}.  All of these color criteria, while defined using different AGN subsamples, are in general agreement with each other, and rely on the fact that AGNs separate cleanly from stars and star-forming galaxies in mid-IR color space~\citep[for a recent empirical discussion of how objects differentiate in \textit{WISE} color space, see][]{Nikutta+14}.  The reason for this separation is because a) stars have nearly blackbody SEDs with flux densities dropping at wavelengths longer than a few microns, and b) while reprocessed photons from dust heating around star formation peaks around a few tens of microns, the hard radiation field from an AGN accretion disk heats dust in the surrounding torus up to the dust sublimation temperature (1,000-1,500~K), leading to a relatively flat power-law spectrum that is easily distinguishable from the aforementioned non-AGN SEDs.  Importantly, because the mid-IR is insensitive to extinction, mid-IR color selection can pick out heavily-obscured or even Compton-thick ($N_\mathrm{H}>10^{24}$~cm$^{-2}$) AGNs \citep{Mateos+13} that are optically indistinguishable from star-forming galaxies \citep[see, for example, Figure 1 in][]{Donley+12}, especially at higher AGN luminosities~\citep{Assef+13,Messias+14,Stern+14}.  This eliminates optical selection effects and can yield much larger and more statistically complete samples of AGNs.

In choosing a mid-IR color selection criteria, we required that the criteria be defined directly from \textit{WISE} data, using only \textit{WISE} data; this rules out the criteria of \citet{Lacy+04}, \citet{Stern+05}, \citet{Lacy+07}, and \citet{Donley+12}, which apply to \textit{Spitzer} IRAC data.  This avoids uncertainties inherent in transforming \textit{Spitzer} magnitudes into \textit{WISE} magnitudes.  This criteria also rules out the SDSS-\textit{WISE} color criteria of~\cite{Wu+12}, which would limit our sample to the SDSS footprint.  We further required that the criteria be defined directly from a clean, highly reliable sample of AGNs and QSOs not defined empirically from \textit{WISE} data; this rules out the criteria of \citet{Jarrett+11}.  Finally, we required that the criteria be a two-color selection, involving $W2$-$W3$, which excludes the \citet{Stern+12} and \citet{Assef+13} criteria.  This is because we are not restricting our sample to high Galactic latitudes, so contamination by brown dwarfs may occur if we only use $W1$-$W2$. We do not know \textit{a priori} that our sources are extragalactic, and some brown dwarfs can share the first color axis $W1$-$W2$ with AGNs due to methane absorption at $3.3\micron$ reducing emission in the $W1$ band \citep[e.g.,][]{Noll+00}.\footnote{See \citet{Kirkpatrick+11, Cushing+11, Mace+13, Cushing+14} for some samples of brown dwarfs discovered with \textit{WISE}.} In developing an all-sky catalog of mid-infrared zero-proper motion, zero-parallax objects, it is key to minimize brown dwarf contamination, as many brown dwarfs are within a few pc and therefore have very high proper motions, as high as $\sim8\arcsec$~yr$^{-1}$ \citep{Luhman+14}.  For picking out \textit{Spitzer} IRAC-selected AGNs down to a limiting magnitude of $W2<15$, the \textit{WISE} one-color AGN selection criteria of \citet{Stern+12} and \citet{Assef+13} are both very reliable, with 95\% and and $>90\%$ reliability, respectively.  However these two-color cuts also overlap significantly with the mid-IR color space occupied by brown dwarfs.  For example, using the representative sample of the AllWISE catalog described in \S\ref{subsec:stellarContamination}, out of 1,056 sources identified as AGNs using the \citet{Stern+12} criterion, 62 (5.9\%) also qualify as brown dwarfs according to the combined empirical criterion for dwarfs with spectral type $\geq$T5 and nearby L and T-type dwarfs of \citet{Kirkpatrick+11}.  Similarly, of the 1,892 sources identified as AGNs using the ``R90'' criterion of \citet{Assef+13}, 59 (3.1\%) fall within the combined \citet{Kirkpatrick+11} criterion.

With these considerations, we chose the two-color selection criterion of \citet{Mateos+12}, who use the Bright Ultrahard \textit{XMM-Newton} Survey (BUXS), one of the largest, flux-limited samples of `ultra hard' (4.5-10 keV) X-ray selected sources, to define a \textit{WISE} AGN selection criterion.  This reduces bias against heavily absorbed AGNs, and the BUXS is comprised mostly (56.2\%) of Type~1 AGNs (emission line widths $>1500$~km~s$^{-1}$); the remainder are almost entirely Type~2 AGNs.  It is well established that the mid-IR luminosities of AGNs correlate strongly with their hard X-ray luminosities \citep[][see also the relation in \citealt{Secrest+15}]{Lutz+04,Gandhi+09,Mateos+15,Stern15}, as Compton up-scattering of UV photons from the accretion disk into the X-ray regime is largely proportional to accretion disk luminosity.  In terms of overlap with the mid-IR color space shared with brown dwarfs, of the 860 sources in our representative sample of the AllWISE catalog that classify as AGNs according to the \citet{Mateos+12} criterion, only 7 (0.8\%) would qualify as L or T dwarfs according to the combined \citet{Kirkpatrick+11} criteria.


We take all sources from the AllWISE catalog following equations (3) and (4) from \citet{Mateos+12} and we require that all of our sources have S/N $\geq$ 5 in the first three bands (\texttt{w1,2,3snr$>=5$}), as recommended in \citet{Mateos+12}, but as a further constraint we limit our results to those with \texttt{cc\_flags} = `0000', meaning that the sources are unaffected by known artifacts such as diffraction spikes, persistence, halos, or optical ghosts.  We subdivided this query into sources above $\delta\geq0\degr$ and source below $\delta<0\degr$ to make our queries tractable.  We concatenated the resultant tables using \textsc{topcat}, version 4.2-3.\footnote{\url{http://www.star.bris.ac.uk/~mbt/topcat/}}  For the remainder of this paper, we refer to AGNs selected in the manner outlined above as mid-IR AGNs (MIRAGNs), although we reiterate that there are other selection criteria for mid-IR AGNs in the literature.

\section{Results}
\label{sec:Results}
\subsection{Sample Properties}
\label{subsec:sampleProperties}
\begin{figure*}
\includegraphics[width=\linewidth]{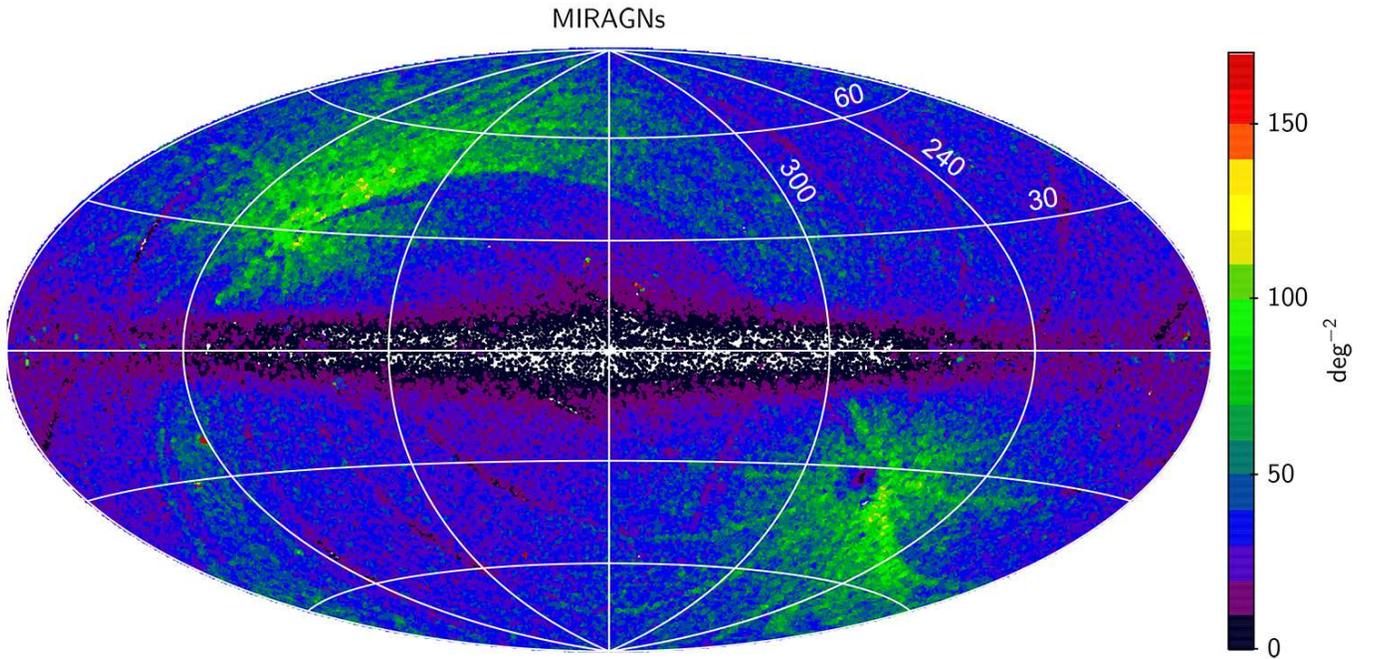}
\caption{Density plot, Aitoff projection in Galactic coordinates, of our full sample of AGN/QSO sources.  The under-density of sources along the Galactic plane below $|b|\lesssim15\deg$ is due to the AGN color criterion effectively excluding stars and other Galactic sources, combined with the source confusion limit of \textit{WISE}~\citep{Wright+10}.  The increased number density at the ecliptic poles is due to deeper \textit{WISE} coverage.}
\label{fig:Aitoffprojection}
\end{figure*}

\begin{figure*}
  \centering
  \subfigure[]{\label{fig:sigSqDegHist}\includegraphics[width=8.7cm]{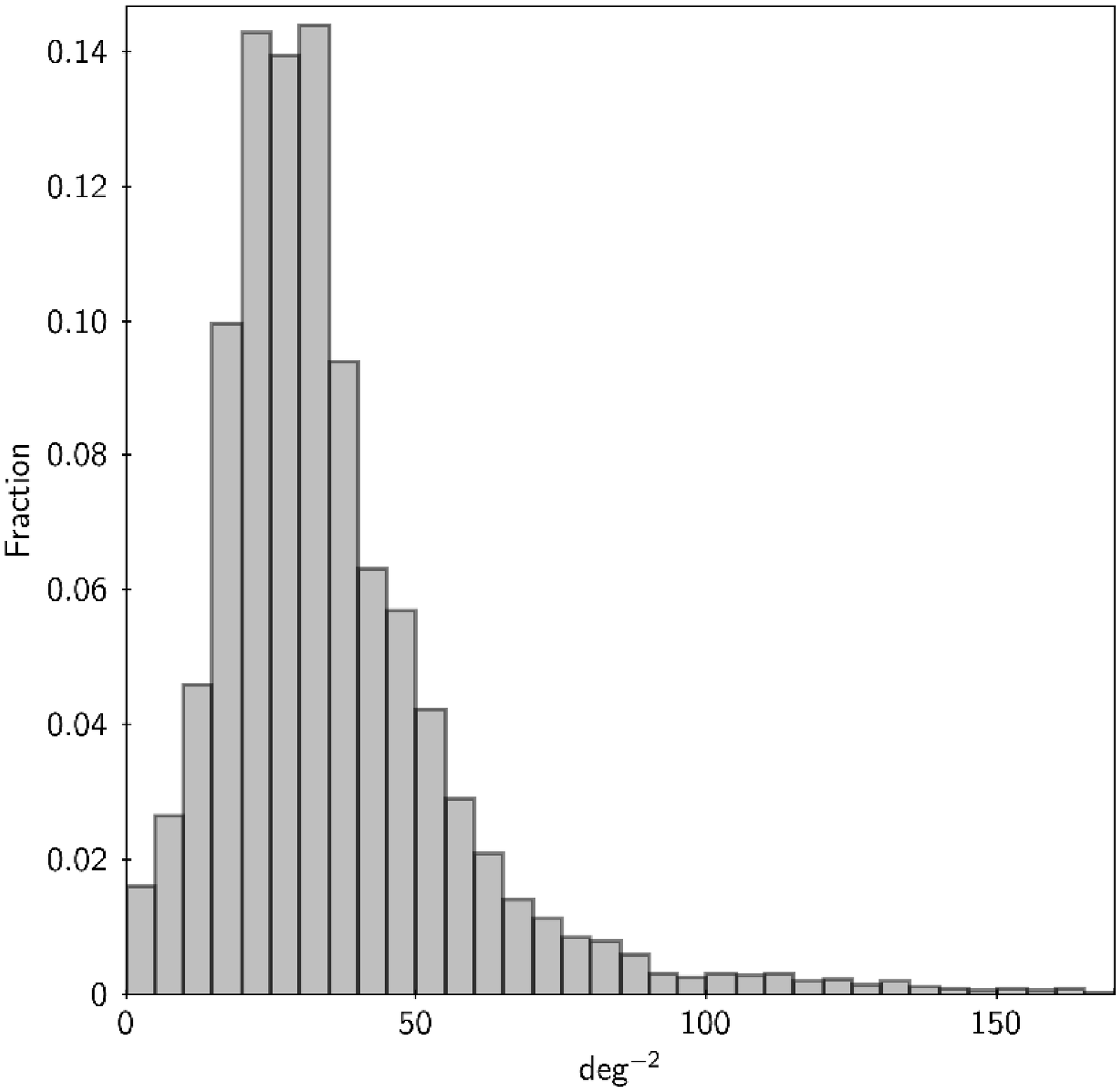}}
  \subfigure[]{\label{fig:sigSqDegCum}\includegraphics[width=8.7cm]{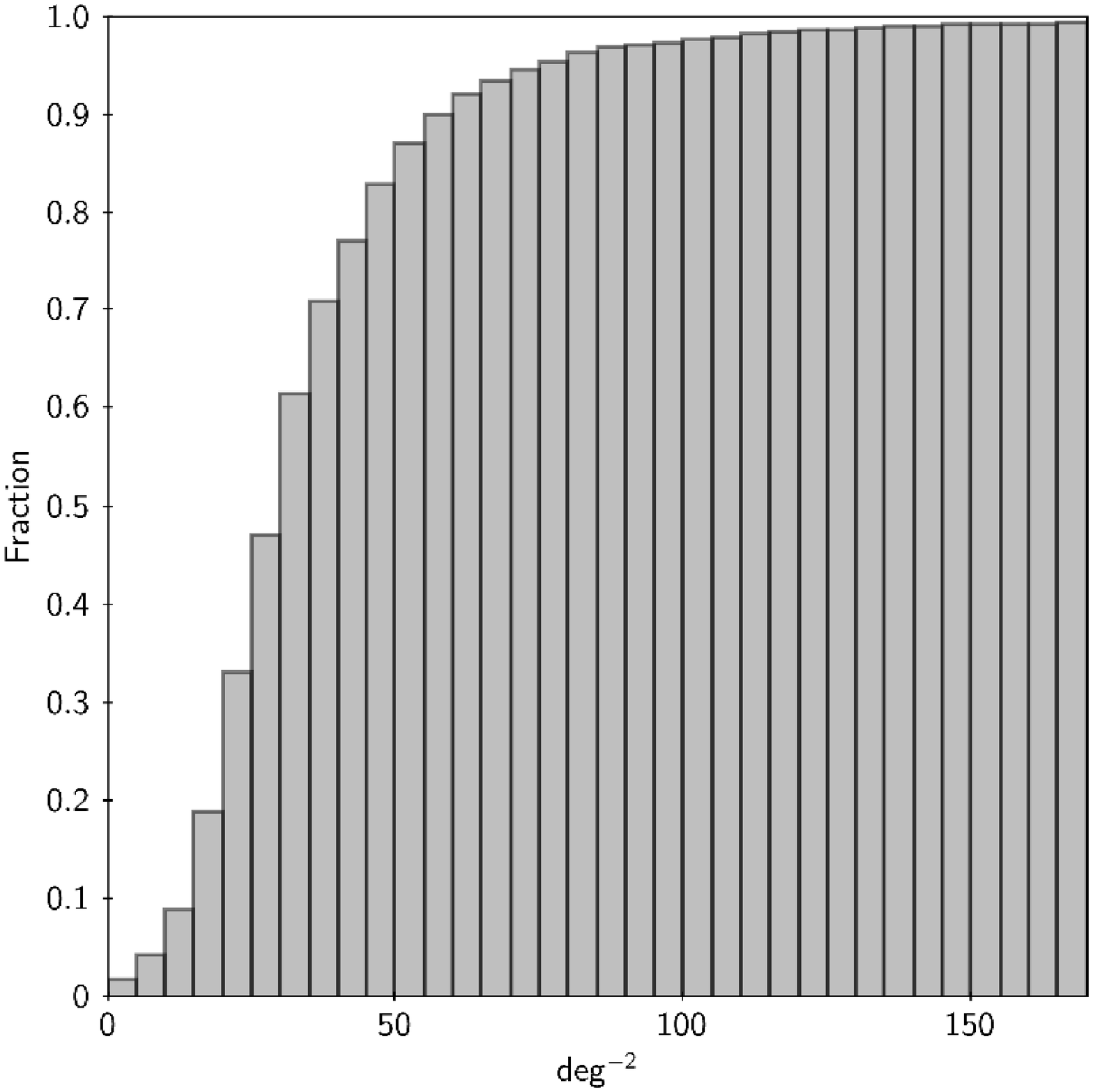}}
  \caption{(a) Normalized histogram of MIRAGNs per deg$^{-2}$; (b) Corresponding cumulative histogram.  The mean source density is $\approx38$~deg$^{-2}$, with 10\% and 90\% thresholds of 15~deg$^{-2}$ and 59~deg$^{-2}$, respectively.\vspace{0.25cm}}
  \label{fig:srcSqDeg}
\end{figure*}

Our sample consists of 1,354,775 MIRAGNs spanning the full sky.  In Figure~\ref{fig:Aitoffprojection}, we show a density plot of MIRAGNs across the sky, clearly showing that our AGN selection criteria is effectively selecting objects outside of the Galaxy.  Our sample is also relatively uniform across the sky.  In Figure~\ref{fig:srcSqDeg}, we show the source density of MIRAGNs.  By randomly sampling 10$^6$ 1-deg$^{2}$ areas across the sky, we calculate a mean source density of $\approx38$~deg$^{-2}$, with 10\% and 90\% thresholds of 15~deg$^{-2}$ and 59~deg$^{-2}$, respectively.  We note that the reasons from the deviation from a truly uniform distribution centered at $N$/$4\pi$\hspace{0.5mm}sr~=~33~deg$^{-2}$ are the over-density of sources at the ecliptic poles (due to deeper \textit{WISE} coverage) and the under-abundance along the Galactic plane (due to source confusion), both effects visible in Figure~\ref{fig:Aitoffprojection}.

\begin{figure*}
\includegraphics[width=\linewidth]{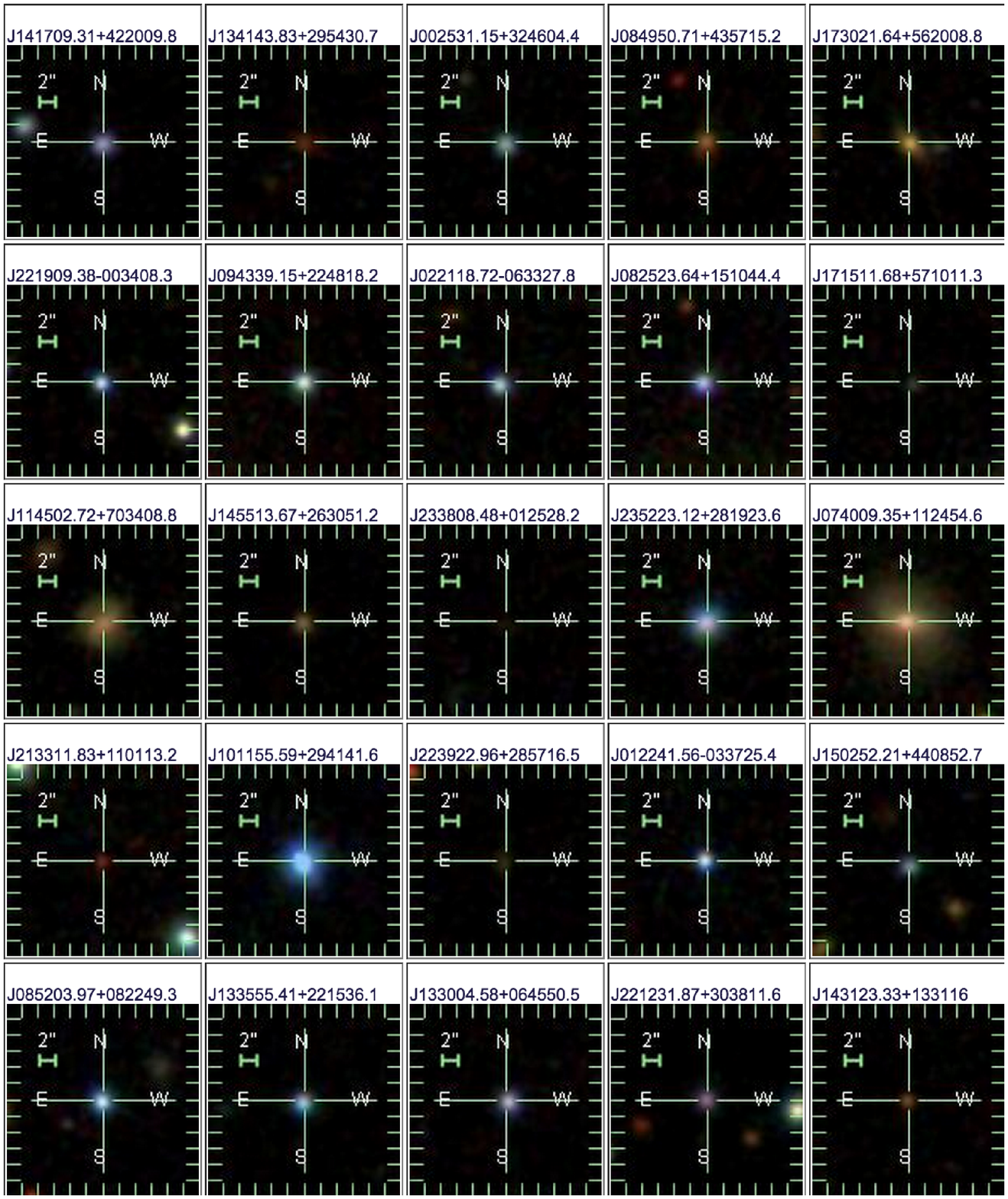}
\caption{Random sample of 25 sources taken from our AllWISE-SDSSDR12 cross match, showing the typical angular extent of sources in our sample. No cuts or photometry requirements were made.}
\label{fig:rand25}
\end{figure*}

Almost none of the MIRAGNs in our sample ($<2\%$) have PSF profile fit $\chi^2_\mathrm{red}>3.0$ in the highest-resolution $W1$ band, indicating that the vast majority of our sources are unresolved by \textit{WISE} and therefore subtend angular scales less than $\sim6\arcsec$.  In Figure~\ref{fig:rand25}, we show SDSS thumbnails of a random selection of 25 sources in our sample, demonstrating that the majority of MIRAGNs in our sample are indeed compact at optical wavelengths as well.

\subsection{Optical Properties}
\label{subsec:sdssdr12}

In order to characterize the optical properties of our sample, we cross-matched it to SDSS-DR12, which is the final data release of SDSS-III~\citep{Eisenstein+11},\footnote{SDSS-DR12 covers 14,555 square degrees, or about 35.3\% of the sky.} within a radial tolerance of $R<1\arcsec$, obtaining 424,366 matches.  To determine the fraction of false positive position matches (that is, incorrectly correlating an object in our sample with a different SDSS object due to random positional agreement), we performed the same match on a scrambled version of our sample coordinates, determining that less than 1\% of our cross-matches are false positive position matches between the two catalogs.

With our list of cross-matches between our sample and SDSS-DR12, we queried the \texttt{PhotoObjAll} table to obtain optical photometric measurements of our sources, requiring the \texttt{clean} photometric quality flag be equal to 1, and obtained 193,637 sources.  We similarly queried the \texttt{SpecObjAll} table to obtain spectroscopic redshift information for our sources, and obtained 39,981 sources.  The following describes the statistics of the sample based on these queries.

\begin{figure}
\includegraphics[width=8.7cm]{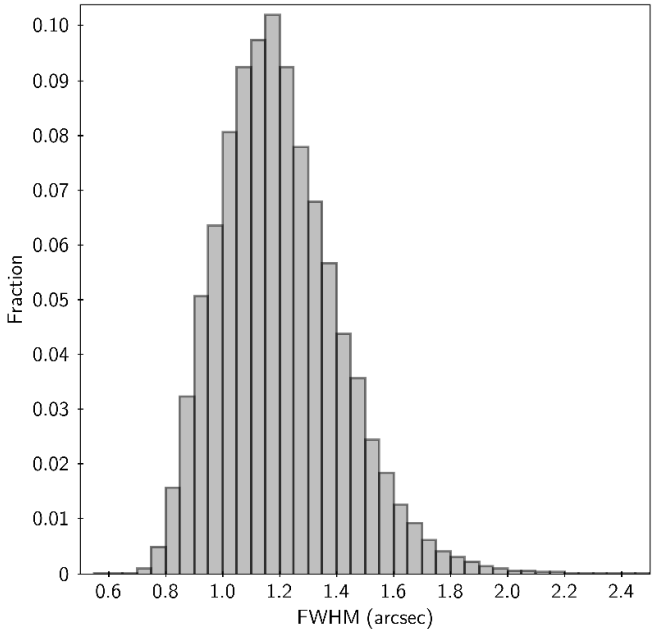}
\caption{Distribution of $r$-band PSF FWHM in our sample cross-matched with SDSS~DR12 \texttt{PhotObjAll}.  Nearly all 193,637 sources in this subsample are unresolved, as we would expect.}
\label{fig:psffwhm}
\end{figure}

\textbf{Source Extent:} In Figure~\ref{fig:psffwhm}, we show the optical extents of sources in our sample, given by the \texttt{psffwhm\_r} parameter.  The majority are not extended, with a mean FWHM of $1.2\pm0.2$~arcsec, comparable with the seeing limit of SDSS~\citep{Stoughton+02}, further underlining the power of our mid-IR selection strategy at picking out unresolved AGNs/QSOs.  

\begin{figure}
\includegraphics[width=8.7cm]{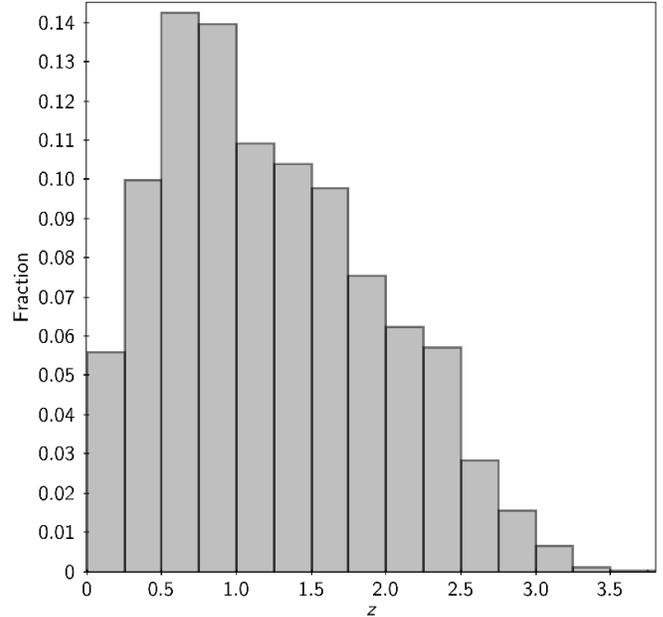}
\caption{Distribution of spectroscopic redshifts of our sample cross-matched with SDSS~DR12 \texttt{SpecObjAll}.}
\label{fig:z}
\end{figure}

\textbf{Distances:} In Figure~\ref{fig:z}, we show the spectroscopic redshifts of MIRAGNs in our sample, which go out to about $z\sim3$.  Note that SDSS spectra come from targeted observations, so this subsample of MIRAGNs suffers from a selection bias and should not be considered to be representative of the physical distribution of redshifts for the entire sample.

\begin{figure}
\includegraphics[width=8.7cm]{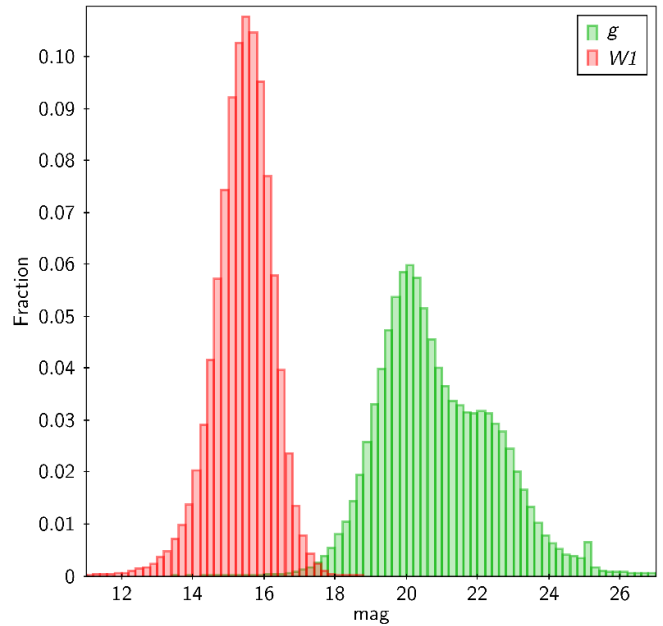}
\caption{Distribution of $W1$ (Vega) and $g$-band (AB) apparent magnitudes.}
\label{fig:mags}
\end{figure}

\textbf{Magnitudes:} In Figure~\ref{fig:mags}, we show the distribution of $W1$ and $g$-band magnitudes for our sources.  In the mid-IR, 1\% of our sources have $W1$ [3.4\micron] magnitudes less than 13; 68\% have $W1$ magnitudes less than 16, and 100\% have $W1$ magnitudes less than 19.  Optically, 2\% of our sources have $g$ magnitudes less than 18; 33\% have $g$ magnitudes less than 20; 75\% have $g$ magnitudes less than 22; and 96\% have $g$ magnitudes less than 24.

Note that the $g$-band magnitudes extend well past the $\sim20$~mag \textit{Gaia} limit.  Extrapolating the number of SDSS DR12-matched sources with $g$-band magnitudes less than 20 to the entire sky, we predict that this sample contains $\sim1.8\times10^5$ of the AGNs/QSOs that will be observed by \textit{Gaia}.

\subsection{Stellar Contamination}
\label{subsec:stellarContamination}

In creating a sample of extragalactic sources without knowledge of their redshifts, it is vital to be able to adequately avoid contamination by stars, especially for unresolved sources.  While stars and AGNs occupy completely different loci in mid-IR color space due to their fundamentally different SEDs \citep[see, for example, Figure 12 in][]{Wright+10}, we nonetheless aim to quantify any possible contamination of our sample by stars.  To do this, we calculate the number of stars in our sample as follows:
\begin{equation}
    N_\mathrm{c,s,AGN} = N \cdot f_\mathrm{c} \cdot f_\mathrm{c,s} \cdot f_\mathrm{c,s,AGN}
\end{equation}
\noindent where $N$ is the size of the AllWISE catalog (747,634,026), $f_\mathrm{c}$ is the fraction of the  AllWISE catalog that has clean \textit{WISE} photometry as per \S\ref{sec:AGNselection} (for the rest of this work, ``clean'' \textit{WISE} photometry means this), $f_\mathrm{c,s}$ is the fraction of those clean sources that are stars, and $f_\mathrm{c,s,AGN}$ is the fraction of stars with clean \textit{WISE} photometry that survive the AGN cut we employ.

\begin{figure}
\includegraphics[width=8.7cm]{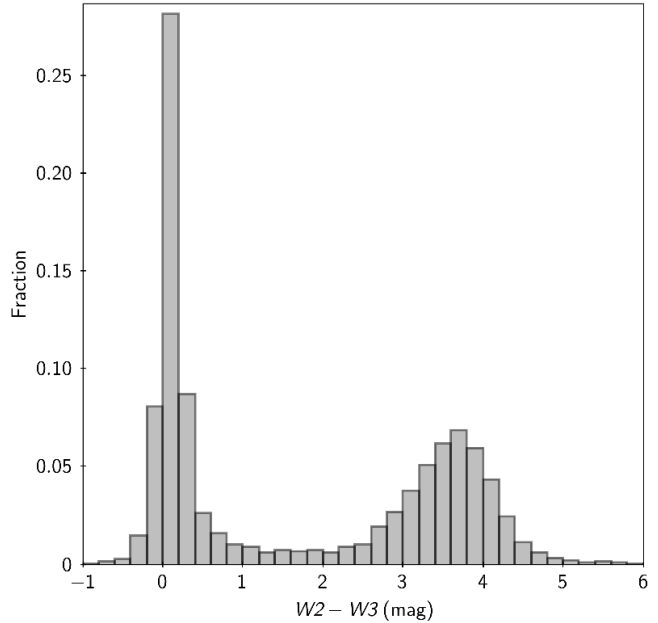}
\caption{Distribution of $W2$-$W3$ for a representative sample of sources from the AllWISE catalog, randomly sampled across the celestial sphere.}
\label{fig:allwisew2w3}
\end{figure}

To estimate $f_\mathrm{c}$, we constructed a representative sample of the AllWISE catalog by querying the AllWISE catalog with a list of 1 million coordinates randomly sampled across the celestial sphere, returning all sources that were within a radius of $R<10''$.  This returned 389,424 unique sources.  Of these, 13,506 have clean \textit{WISE} photometry, or 3.47\%.  Estimating $f_\mathrm{c,s}$ can be done by utilizing the fact that stars cleanly separate from extragalactic objects in mid-IR color space, especially in $W2$-$W3$.  This is shown more explicitly in Figure~\ref{fig:allwisew2w3}, which depicts a clearly bimodal distribution of stars and extragalactic objects in $W2$-$W3$.  We select all objects with clean \textit{WISE} photometry and with $W2$-$W3<2$ as ``stars'', although there is some mixing of stars and extragalactic objects between $W2$-$W3\approx$~1-2, and elliptical galaxies have star-like mid-IR colors, so this selection should be viewed as conservative.  From this, we estimate that approximately $55.6\%$ of the sources in the AllWISE catalog with clean photometry are stars.

Estimating the value of $f_\mathrm{c,s,AGN}$ can be done by constructing an unambiguous sample of stars and then cross-matching the sample with the AllWISE catalog.  We created such a sample of stars from the all-sky PPMXL catalog \citep{Roeser+10}, a catalog of $\sim900$~million sources with optical photometric and astrometric information from USNO-B1.0 and 2MASS, which is complete down to approximately $V\approx20$. We selected all sources in the PPMXL catalog with $B$-band magnitude less than 12, as this excludes all AGNs and most extragalactic objects in the PPMXL catalog.\footnote{As a quality control, we required that the difference in $B$ magnitude between the first and second epochs of USNO-B1.0 be less than 0.5 mag.}  We further required that the absolute values of the proper motions in RA and Dec be less than 150 mas~yr$^{-1}$, as this avoids spurious entries in the PPMXL catalog.  We then cross-matched this sample of stars to the AllWISE catalog within $R<1\arcsec$, returning 499,724 sources with clean \textit{WISE} photometry.  Of these, only 20 (0.0040\%) fall within our AGN selection criterion.\footnote{Because of the brighter magnitudes of our star sample, saturation effects in the corresponding AllWISE data become important.  To avoid stars with spurious AllWISE magnitudes, we removed any with $W1<8$, $W2<7$, and $W3<3.8$.  This removed 23 stars that would have been classified as AGNs according to our criterion.  This is not a significant effect for our sample of AGNs, however, affecting less than 0.092\% of our sample.}  To estimate the number of sources that would have fallen within our AGN criterion by chance mismatch, we cross-matched a scrambled version of our list of star coordinates to the AllWISE catalog, returning 3,626 ``matches'' within $R<1\arcsec$.  Of these, 7 fall within our AGN selection criterion, implying that many, if not most of the 20 stars falling within our AGN criterion are actually chance mismatches.  We therefore consider 0.0040\% to be an upper limit to the percentage of stars with \textit{WISE} colors following our criterion.


Multiplying these fractions together, $N_\mathrm{c,s,AGN}\approx 580$.  Given our conservative estimates described above, we interpret this as the upper limit of the total number of AllWISE-observed stars that leak into our sample.  With $1.4\times10^6$ sources in our MIRAGN sample, the expected contamination of the total sample by stars is therefore $\le0.041\%$.  We conclude that nearly 100\% of our sample is uncontaminated by stars. We term this the ``reliability'' of our sample.  It is important to note that the types of stars found in PPMXL, an optical catalog, are by no means the same types of stars found in AllWISE and many infrared-bright stars, such as L and T-type dwarfs, are under-represented in our star sample.  We have chosen a two-color mid-IR AGN selection criterion partly to avoid contamination by brown dwarfs, but we emphasize that our reliability analysis pertains specifically to optical survey work.

\subsection{Completeness}
\label{subsec:completeness}

\begin{figure*}
  \centering
  \subfigure[]{\label{fig:lqaq2midir}\includegraphics[width=8.7cm]{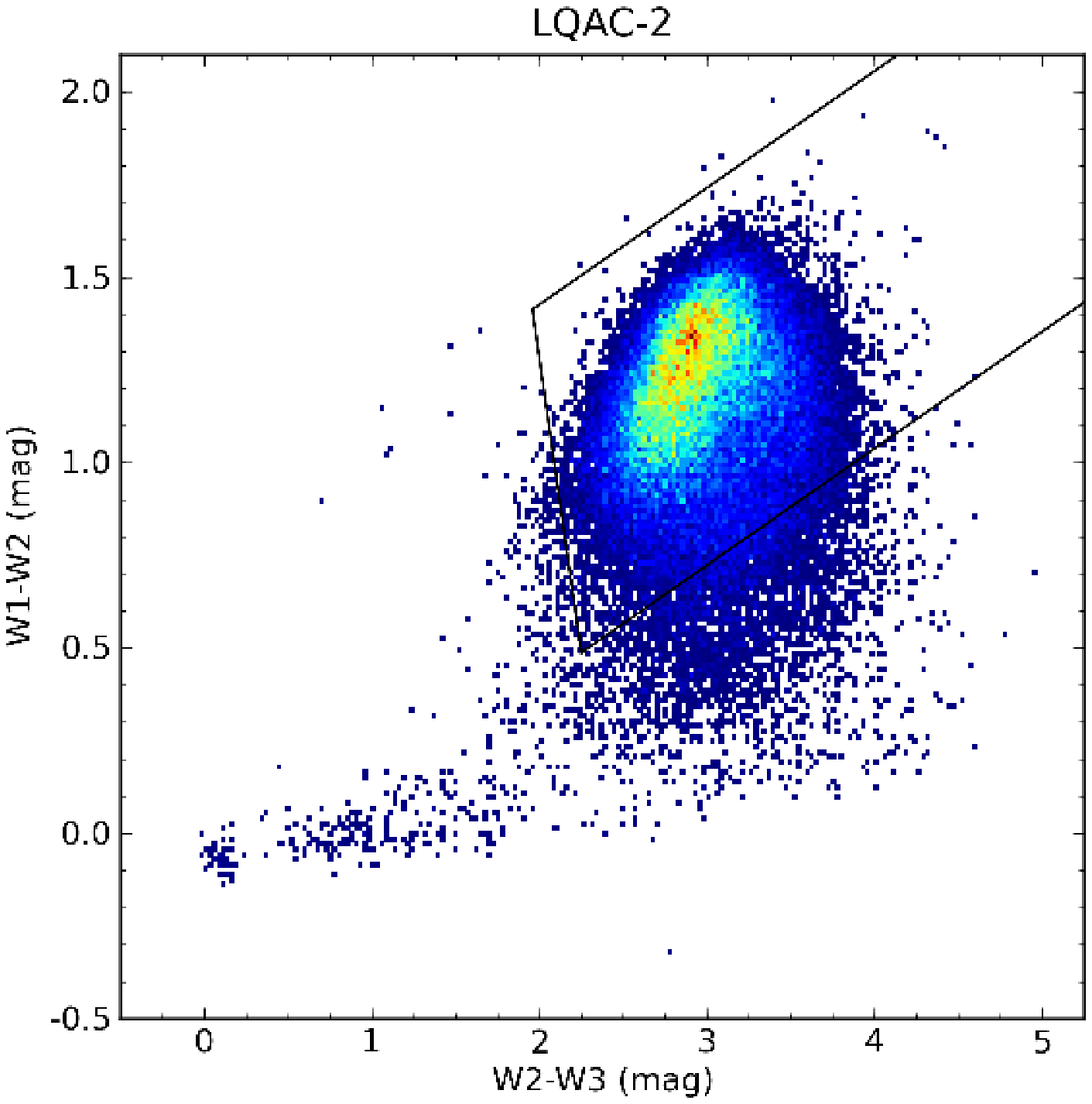}}
  \subfigure[]{\label{fig:dr12qmidir}\includegraphics[width=8.7cm]{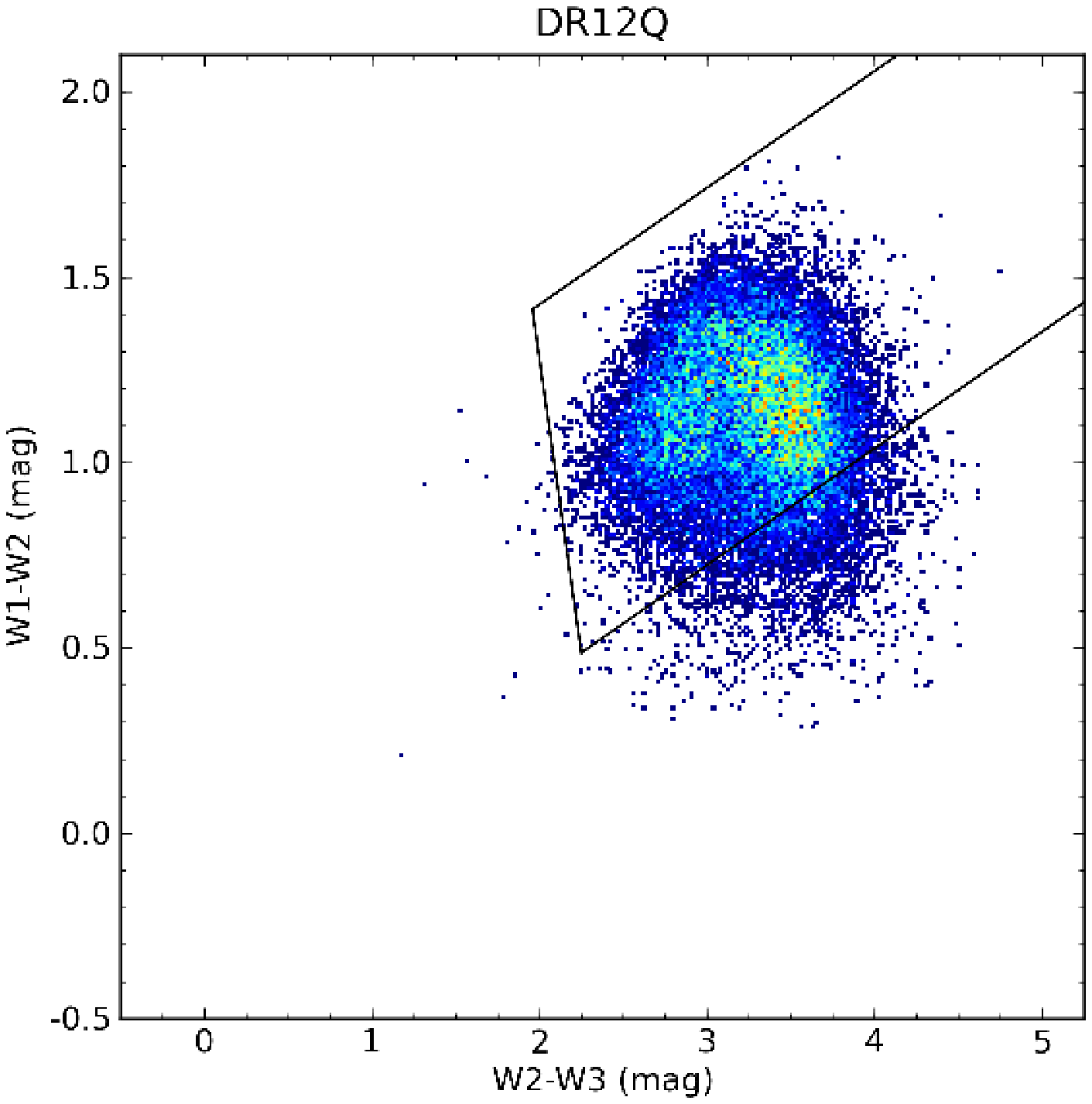}}
  \caption{(a) Mid-IR color-color plot for $R<19$ sources in LQAC-2; (b) For DR12Q with $g<20$.  The black lines are the~\citet{Mateos+12} demarcation. (See \S\ref{subsec:completeness})\vspace{0.25cm}}
  \label{fig:lqac2dr12qmidir}
\end{figure*}

While the primary objective of this study is to obtain a \textit{reliable} sample of extragalactic sources using mid-IR color-selected AGNs, it is nonetheless useful to explore the statistical completeness of our sample.  To do this, we used the second release of the Large Quasar Astrometric Catalog~\citep[LQAC-2;][]{Souchay+12}, which contains 187,504 quasars, including radio-selected quasars from the ICRF2, optically-selected quasars from SDSS, and infrared-selected quasars from 2MASS; and so thus represents a robust sample of quasars over a wide range of wavelengths.

After cross-matching with AllWISE, we find that 93,403 quasars from LQAC-2 have clean detections.  The majority of non-detections are due to sources in LQAC-2 that are too faint, having $R\gtrsim19$.\footnote{166,033 quasars in LQAC-2 have $R$-band magnitudes in the catalog, and 104,656 (63.03\%) have $R\geq19$.}  Of the 61,377 sources in LQAC-2 brighter than this limit, 51,618 (84.10\%) have clean detections with \textit{WISE}.  Of these, 46,928 are MIRAGNs, or 90.91\%.  Broken down by wavelength-based source catalog \citep[see Table 1 in][]{Souchay+12}:\\ \\
radio:  \hfill 84.00\% (15,441/18,391)\\
near-IR: \hfill 82.47\% (15,973/19,368)\\
optical:\footnote{The difference of 1 between this denominator and the full sample is due to the $R$ magnitudes deriving from complementary USNO-B1.0~\citep{Monet+13} data, so one source in LQAC-2 is a purely radio-determined AGN with a cross-identification in USNO-B1.0: most sources in LQAC-2 have data from multiple wavelengths.} \hfill 90.91\% (46,927/51,617)\\ \\

\noindent With a magnitude-limited sample, mid-IR AGN classification is therefore quite complete, even for AGNs selected from radio surveys.  Finally, we note that of the 3,414 ICRF2 sources, 1,364 have clean \textit{WISE} detections and $R<19$.  Of these, 1,219 (89.40\%) are MIRAGNs.

It is of interest to compare the completeness of MIRAGNs with the number of quasars expected to be discovered by \textit{Gaia} down to a magnitude of $V$~=~20 \citep[$\approx5\times10^5$,][]{Mignard+12}.  To do this, we performed a similar analysis using the DR12Q catalog from SDSS, which contains 297,301 quasars, 44,831 of which have clean \textit{WISE} detections.  The majority of non-detections is again due to a limiting magnitude of about $g<20$.  Of those clean detections with $g<20$, 38,915 (86.8\%) are MIRAGNs.  Using the $g$-band as a proxy for $V$, 23,906/27,093 (88.2\%) of cleanly-detected sources with $g<20$ are MIRAGNs.  Extrapolating over the whole sky (the BOSS survey covers $\approx9.2\times10^3$~deg$^2$), $\approx8.3\times10^4$ MIRAGNs in our sample with $g<20$ are outside the SDSS footprint and are therefore expected to be new.

\begin{figure}
\includegraphics[width=8.7cm]{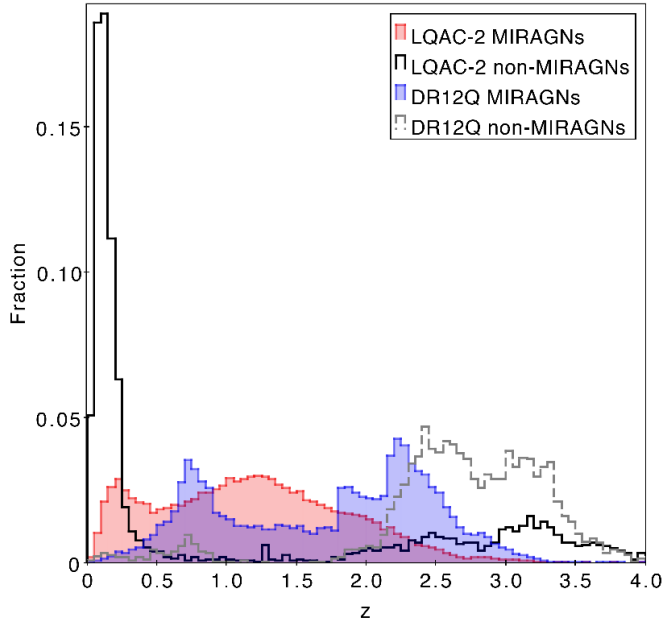}
\caption{Distribution of redshifts for AGNs in the LQAC-2 and DR12Q catalogs meeting the \textit{WISE} photometry requirements outlined in \S\ref{sec:AGNselection}, and brighter than $R<19$ and $g<20$, respectively.\vspace{0.25cm}}
\label{fig:LQAC2-DR12Q-zvsAGNhisto}
\end{figure}

It is worth discussing the reasons why $\approx10\%-20\%$ of AGNs in LQAC-2 and DR12Q within our magnitude limit are not recovered in our sample of MIRAGNs.  In Figure~\ref{fig:lqac2dr12qmidir} we show the mid-IR color distribution of (a) the LQAC-2 catalog, and (b) the DR12Q catalog.  Most sources that do not meet the criteria are bluer in their $W1$-$W2$ colors, which is due to two effects.  First, lower ratios of AGN/host galaxy luminosities preclude inclusion by our criteria.  For example, \citet{Mateos+12} found that, above a hard X-ray luminosity of $L_\mathrm{2-10~keV}\geq10^{44}$~erg~s$^{-1}$, 97.1\% of type 1 AGNs (emission line widths $\geq1500$~km~s$^{-1}$) and 76.5\% of type 2 AGNs (emission line widths $<1500$~km~s$^{-1}$) fall within their criteria.  Below $L_\mathrm{2-10~keV}<10^{44}$~erg~s$^{-1}$, those percentages are 84.4\% and 39.1\%, respectively, and contamination by the host galaxy becomes more evident.  At lower redshifts ($z<0.5$), a larger fraction of AGNs in both DR12Q but especially LQAC-2 have bluer $W1$-$W2$ colors that fall outside the \citet{Mateos+12} criteria (Figure~\ref{fig:LQAC2-DR12Q-zvsAGNhisto}).  To show that this is due to lower AGN/host galaxy luminosities, we calculated the rest-frame 5~GHz luminosities of AGNs with radio data in LQAC-2 by assuming power-law radio SEDs of the form $f_\nu\propto\nu^{-\alpha}$, and calculating $\alpha$ directly from the catalog 2~GHz and 5~GHz spectral energy densities.  On average, AGNs in LQAC-2 have power law indices of $\alpha=0.3\pm0.6$, in line with the flat-spectrum radio SEDs typically seen in AGNs.  Using $\alpha$ to correct for redshift, we find that for sources below a redshift of $z<0.5$ the mean 5~GHz luminosity is $L_\mathrm{5\hspace{0.5mm}GHz}=2.5\times10^{41}$~erg~s$^{-1}$, while for sources above a redshift of $z>0.5$ the mean 5~GHz luminosity is $L_\mathrm{5\hspace{0.5mm}GHz}=7.9\times10^{43}$~erg~s$^{-1}$.  AGNs at higher redshift in LQAC-2 are therefore much more dominant, and so tend to manifest as MIRAGNs.\footnote{Although more luminous AGNs may also be preferentially found in more luminous host galaxies.  See, for example~\citet{Hamilton+08}.}

Second, at very high redshifts ($z>2$), the SEDs of even pure AGNs begin to move out of the mid-IR color-color demarcation~\citep[see Figure 5 in][]{Mateos+12}.  This effect can be seen in Figure~\ref{fig:LQAC2-DR12Q-zvsAGNhisto}, and affects the DR12Q catalog especially due to the higher average redshift ($z=2.1$) of AGNs in the catalog.

\begin{deluxetable*}{ccccccccc}

\tablecolumns{9}
\tablecaption{
MIRAGN Data
\label{tab:MIRAGNdata}
}
\tablehead{
\colhead{AllWISE} & \colhead{RA} & \colhead{Dec} & \colhead{$W1$-$W2$} & \colhead{$W2$-$W3$} & \colhead{$W1$} & \colhead{$g$} & \colhead{$z$} & \colhead{$z$-type} 
}

\startdata
J005947.42-011010.8 &  14.947587 &  -1.169694 &  1.26 &  2.74 &  14.73 &  18.82 &  1.123 &  s  \\
J005939.81-011445.7 &  14.915884 &  -1.246029 &  0.73 &  2.99 &  14.81 &  21.35 &  0.457 &  s  \\
J010016.47-020912.8 &  15.068651 &  -2.153577 &  1.31 &  3.57 &  15.15 &    &  2.009 &  s  \\
J005858.94-010459.4 &  14.745614 &  -1.083188 &  1.01 &  2.92 &  15.12 &    &   &       \\
J005847.49-010549.7 &  14.697876 &  -1.097150 &  0.71 &  2.76 &  12.26 &    &  0.047 &  s  \\
J010210.49-015630.0 &  15.543743 &  -1.941685 &  0.92 &  2.38 &  14.74 &    &  0.6 &  p  \\
J010323.03-012637.0 &  15.846000 &  -1.443626 &  1.19 &  3.19 &  14.79 &    &   &      \\
J010032.21-020046.0 &  15.134236 &  -2.012790 &  1.11 &  3.35 &  12.86 &    &  0.227 &  s  \\
J005847.77-020808.6 &  14.699062 &  -2.135742 &  1.27 &  2.89 &  15.48 &    &   &      \\
J005936.26-013003.8 &  14.901093 &  -1.501077 &  1.14 &  3.51 &  15.17 &  19.46 &  1.016 &  s  \\
J005814.49-011507.0 &  14.560381 &  -1.251954 &  0.99 &  2.61 &  14.39 &    &   &       \\
J010220.08-005743.5 &  15.583695 &  -0.962098 &  1.31 &  2.95 &  15.07 &    &   &       \\
J005804.98-015015.7 &  14.520752 &  -1.837710 &  0.88 &  3.37 &  13.84 &    &  0.239 &  s  \\
J010246.00-012600.5 &  15.691669 &  -1.433499 &  1.38 &  3.50 &  15.31 &    &   &       \\
J010248.37-021532.9 &  15.701556 &  -2.259148 &  1.65 &  3.77 &  15.38 &    &   &       \\
J010304.14-010040.1 &  15.767260 &  -1.011162 &  0.90 &  2.89 &  15.24 &    &   &       \\
J010114.58-021141.1 &  15.310782 &  -2.194774 &  1.02 &  3.70 &  16.04 &    &   &       \\
J010010.93-014909.5 &  15.045565 &  -1.819306 &  1.05 &  3.43 &  15.66 &    &   &       \\
J010003.46-015427.7 &  15.014454 &  -1.907698 &  1.58 &  4.19 &  17.25 &    &   &      \\
J010249.02-010545.0 &  15.704260 &  -1.095849 &  1.20 &  2.68 &  15.01 &  19.89 &  1.038 &  s  \\
J005956.19-010722.6 &  14.984129 &  -1.122969 &  1.13 &  3.74 &  15.94 &    &   &       \\
J005957.93-005311.4 &  14.991392 &  -0.886508 &  1.46 &  3.37 &  15.85 &    &   &      \\
J010129.88-010515.9 &  15.374504 &  -1.087752 &  1.62 &  3.22 &  15.15 &    &   &       \\
J010104.16-005918.6 &  15.267358 &  -0.988513 &  1.40 &  3.71 &  15.84 &    &   &       \\
J005939.26-014846.1 &  14.913610 &  -1.812812 &  1.41 &  2.99 &  14.98 &    &   &       
\enddata

\tablecomments{``AllWISE" is the ``designation'' column in the AllWISE catalog; RA and Dec are the AllWISE catalog coordinates, in J2000. $g$-band magnitudes come from LQAC-2 where available, DR12Q else.  Redshifts come preferentially from LQAC-2, then the non-flagged pipeline redshifts from DR12Q, then MILLIQUAS.  Photometric redshifts from MILLIQUAS are flagged in $z$-type as `p'.   The remaining columns are the unique string identifiers found in the matched catalogs, but have been omitted from this printing for space.  The full version of this table will be made available online.\vspace{0.25cm}}
\end{deluxetable*}

If we exclude sources in the LQAC-2 with $z>2$ and $R\geq19$, 93.0\% of the remaining sources are MIRAGNs.  If we make the same redshift cutoff for sources in DR12Q with $g\geq20$, 98.2\% of the remaining sources are MIRAGNs.  We note that the many (32.7\% and 51.0\%, respectively) of the high redshift sources excluded in our sample would have been included if we had used the one-color criteria ($W1$-$W2>0.8$) of ~\citet{Stern+12}.  The AGN criteria outlined in \S\ref{sec:AGNselection} thus minimizes leakage of stars into our sample at the cost of some missed AGNs (i.e., false negatives, or ``type II errors'').

Finally, \citet{DiPompeo+15} recently published a catalog of over 5.5~million quasar candidates with SDSS+\textit{WISE} photometry.  Extending the \textit{XDQSOz} model of \citet{Bovy+12} to include \textit{WISE} photometry, and using a training set of spectroscopically confirmed quasars, they assigned quasar probabilities $P_\textrm{QSO}$ for unresolved sources from SDSS~DR8, as well as computing photometric redshifts, which they find to be significantly improved by the addition of \textit{WISE} photometry.  We cross-matched our sample to their catalog to within $R<1\arcsec$, as before, finding 227,011 matches. We found several reasons for non-matches, which we outline below.

The first and most significant reason for non-matches with our sample is the limiting magnitude.  Of the $\sim3.7$~million sources in their catalog with high quality SDSS photometry (PSF $g$ S/N~$\geq$~5), only $\sim9\%$ have $g<20$.\footnote{These are Pogson magnitudes we calculated as explained in \url{https://www.sdss3.org/dr8/algorithms/magnitudes.php}.} The second reason for non-matches is the differing photometry requirements between our sample and the \textit{XDQSOz} catalog.  Of the 197,635 sources in the \textit{XDQSOz} catalog with $g<20$ not in our sample, 91,711 are not in the AllWISE catalog at all, likely due to the independent ``forced photometry'' performed on \textit{WISE} data for the \textit{XDQSOz}.\footnote{Within a $1\arcsec$ radius; relaxing the radius to $3\arcsec$ yields an additional 14,476 candidate sources in the AllWISE catalog.}   Of the sources not in our sample that are in the AllWISE catalog, only 5,993 fulfill our photometry quality requirements as outlined in \S\ref{sec:AGNselection}.  For the sources in \textit{XDQSOz} that are in the AllWISE catalog with clean \textit{WISE} photometry brighter than $g<20$, 130,420/136,413 (95.6\%) are in our sample.  Finally, a third reason for non-matches is again the redshift limitation of our chosen AGN selection criterion of $z\sim2$.  If we retain only sources with $z<2$, $g<20$, and with clean \textit{WISE} photometry, 110,756/112,138 (98.8\%) of sources in \textit{XDQSOz} are in our sample.  We note that relaxing our match radius to $R<3\arcsec$ does not significantly alter our results, yielding a total of 236,963 matches between our sample and the \textit{XDQSOz} catalog, with 98.4\% of sources in \textit{XDQSOz} with $z<2$, $g<20$, and with clean \textit{WISE} photometry in our sample.



\subsection{New AGNs}
To estimate how many sources in our sample are expected to be previously uncatalogued AGNs, we cross-matched our sample to the Million Quasars (MILLIQUAS) Catalog, version 4.5\footnote{\url{http://quasars.org/milliquas.htm}}, a heterogenous compilation of all known or candidate AGNs/QSOs through May 2015, which contains 1,153,110 entries.\footnote{MILLIQUAS includes the Half Million Quasars (HMQ) Catalogue \citep{Flesch15}, but the latter excludes \textit{candidate} AGNs/QSOs.  Of AGNs in the HMQ, 55.33\% of sources with $R<19$ are MIRAGNs, suggesting a prevalence of non-dominant, low-redshift sources or sources at high redshift.  Indeed, by further excluding sources in the HMQ with $z<0.5$ and $z>2.0$ (see \S\ref{subsec:completeness}), 79.19\% of sources in the HMQ are MIRAGNs.}   It includes quasar data from the NASA/IPAC Extragalactic Database (NED), SIMBAD, and SDSS-DR12Q.  Cross-matching our source list to the MILLIQUAS catalog to within $R<1\arcsec$ yields 202,203 sources, however, in order to more completely assess the number of sources in our sample already in the MILLIQUAS catalog, we relaxed our cross-match radius to $R<10\arcsec$, obtaining 210,534 matches.  To estimate the level of contamination at this cross-match radius by random matches, we cross-matched a scrambled version of our sample coordinates with MILLIQUAS, obtaining 976 matches within $R<10\arcsec$, a contamination level of 0.46\%.  Expanding our match radius to $R<30\arcsec$ only produces an additional 6,216 matches, and an unacceptable level of contamination by random matches of 4.0\%.  Our sample of mid-IR selected AGNs is therefore expected to contain approximately 1.1~million uncatalogued AGNs.  We give an example of our sample, cross-matched to the LQAC-2, DR12Q, and MILLIQUAS catalogs to within $R<1\arcsec$, in Table~\ref{tab:MIRAGNdata}.

\section{Conclusions}
\label{sec:Conclusions}

We have explored the use of \textit{WISE}-selected AGNs to derive a highly reliable sample of extragalactic sources for astrometric purposes.  Our primary conclusions are as follows:
\begin{enumerate}
   \item Using the two-color AGN criteria of~\citet{Mateos+12} and strict photometric quality requirements, we derive a sample of $1,354,775$ mid-IR AGNs from the AllWISE source catalog.  Approximately 1.1~million of these were previously uncatalogued.  
   \item From a reliability analysis using 499,724 stars from the PPMXL catalog, we estimate that the fraction of stars in our sample is extremely small, $\le0.041\%$, and conclude that this technique is extremely reliable.
   \item The use of mid-IR color selection results in a high level of completeness, and we estimate that our sample contains $\approx8.3\times10^4$ AGNs/QSOs with $g$-band magnitudes below $<20$ that fall outside the SDSS footprint, a significant fraction ($\approx17\%$) of the number of QSOs expected to be discovered by \textit{Gaia}.
\end{enumerate}

In a subsequent paper, we will use the sample derived here to look for optical signatures of previously undetected AGN using multi-epoch URAT observations. URAT is an all-sky astrometric survey in the visible, and is a follow-up project to the previous UCAC program. Utilizing the red-lens from the UCAC program, the telescope has been completely redesigned. The new 4-shooter camera consist of four large 10,560 by 10,560 pixel CCDs, with a combined single exposure covering 28 square degrees of the sky at a resolution of $0.9\arcsec$~pix$^{-1}$. The newly released URAT1 catalog \citep{Zacharias+15,Zacharias15} contains accurate positions (typically 10 to 30 mas std.~error) of 220 million stars in the 3 to 18.5 magnitude range, mainly in the northern hemisphere. Proper motions have been obtained for 85\% of these stars utilizing the Two Micron All Sky Survey (2MASS) as first epoch. URAT1 is also supplemented by 2MASS and AAVSO Photometric All-Sky Survey (APASS) photometry. We will characterize the astrometric and photometric variability of the AGNs we detect with URAT, we will provide an optical catalog of these objects and their derived properties, and we will explore the utility of sources identified in this paper for future ICRF work.


\acknowledgements

We thank the anonymous referee for their very thorough review that significantly improved the clarity of this work.  We also thank Daniel Stern (JPL/Caltech) for his helpful discussion on this project, and we thank Ciprian Berghea (USNO) for his help with Pan-STARRS.

This publication makes use of data products from the \textit{Wide-field Infrared Survey Explorer}, which is a joint project of the University of California, Los Angeles, and the Jet Propulsion Laboratory/California Institute of Technology, and NEOWISE, which is a project of the Jet Propulsion Laboratory/California Institute of Technology. \textit{WISE} and NEOWISE are funded by the National Aeronautics and Space Administration.

Funding for SDSS-III has been provided by the Alfred P. Sloan Foundation, the Participating Institutions, the National Science Foundation, and the U.S. Department of Energy Office of Science. The SDSS-III web site is http://www.sdss3.org/. SDSS-III is managed by the Astrophysical Research Consortium for the Participating Institutions of the SDSS-III Collaboration including the University of Arizona, the Brazilian Participation Group, Brookhaven National Laboratory, Carnegie Mellon University, University of Florida, the French Participation Group, the German Participation Group, Harvard University, the Instituto de Astrofisica de Canarias, the Michigan State/Notre Dame/JINA Participation Group, Johns Hopkins University, Lawrence Berkeley National Laboratory, Max Planck Institute for Astrophysics, Max Planck Institute for Extraterrestrial Physics, New Mexico State University, New York University, Ohio State University, Pennsylvania State University, University of Portsmouth, Princeton University, the Spanish Participation Group, University of Tokyo, University of Utah, Vanderbilt University, University of Virginia, University of Washington, and Yale University.


\end{document}